\documentclass{webofc}
\usepackage[varg]{txfonts}   
\begin{document}
\title{Conformality and percolation threshold in neutron stars}
%
%

\author{
\firstname{Michał} \lastname{Marczenko}\inst{1}\fnsep\thanks{\email{michal.marczenko@uwr.edu.pl}} \and
\firstname{Larry} \lastname{McLerran}\inst{2} \and
\firstname{Krzysztof} \lastname{Redlich}\inst{3} \and
\firstname{Chihiro} \lastname{Sasaki}\inst{3}}

\institute{Incubator of Scientific Excellence - Centre for Simulations of Superdense Fluids, University of Wroc\l{}aw, plac Maksa Borna 9, 50-204 Wroc\l{}aw, Poland 
\and
Institute for Nuclear Theory, University of Washington, Box 351550, Seattle, Washington 98195, USA
\and
Institute of Theoretical Physics, University of Wroc\l{}aw, plac Maksa Borna 9, 50-204 Wroc\l{}aw, Poland}

\abstract{%
Speed of sound is given attention in multi-messenger astronomy as it encodes information of the dense matter equation of state. Recently the trace anomaly was proposed as a more informative quantity. In this work, we statistically determine the speed of sound and trace anomaly and show that they are driven to their conformal values at the centers of maximally massive neutron stars. We show that the local peak in the speed of sound can be associated with deconfinement along with percolation conditions in QCD matter.}
\maketitle

\section{Introduction}


The thermodynamic properties of matter are encoded in the speed of sound. In the context of dense nuclear matter, it provides insight into the structure of the equation of state (EoS) at densities beyond the saturation density ($n_{\rm sat}=0.16~\rm fm^{-3}$)~(see, e.g.,~\cite{Tan:2021nat, Zhang:2019udy, Tews:2018kmu, Legred:2021hdx}). Such high densities can be found in neutron stars (NSs), which makes them unique extraterrestrial laboratories for probing dense matter. Unfortunately, the density range found in NSs is not accessible by first-principle methods and the structure of the EoS remains unknown. Nevertheless, great progress in constraining the EoS was achieved by systematic analyses of recent astrophysical observations of the massive pulsar PSRJ0740+6620~\cite{Cromartie:2019kug, Fonseca:2021wxt, Miller:2021qha, Riley:2021pdl} and PSR J0030+0451~\cite{Miller:2019cac} by the NICER collaboration, and the constraint from the GW170817 event~\cite{LIGOScientific:2018cki}, within parametric models of the EoS~(see, e.g.,~\cite{Alford:2013aca, Alford:2017qgh, Annala:2019puf, Annala:2017llu}). At zero temperature, the speed of sound is expressed as 
\begin{equation}
c_s^2 = \frac{n_B}{\mu_B\chi_B}\textrm,
\end{equation}
where
\begin{equation}
\chi_B = \frac{\partial n_B} {\partial \mu_B}
\end{equation}
is the susceptibility of the net-baryon number density. Recently, it has been argued that at low temperatures and large baryon densities, cumulants of the net-baryon number measured in heavy-ion collision experiments~\cite{Bazavov:2012vg, Borsanyi:2014ewa, Karsch:2010ck, Braun-Munzinger:2014lba, Vovchenko:2020tsr, Friman:2011pf} may allow for direct measurement of the speed of sound~\cite{Sorensen:2021zme}. They are expected to probe the critical behavior at the QCD phase boundary~\cite{Stephanov:1999zu, Asakawa:2000wh, Hatta:2003wn} and the QCD critical point in the beam energy scan programs at the Relativistic Heavy Ion Collider at Brookhaven National Laboratory and the Super Proton Synchrotron at CERN.

The trace anomaly scaled by the energy density was recently proposed as a measure of conformality~\cite{Fujimoto:2022ohj}
\begin{equation}
\Delta = \frac{1}{3} - \frac{p}{\epsilon} \textrm.
\end{equation}
As the scale invariance is expected to be restored at asymptotically high temperature/density in QCD, $\Delta\rightarrow0$ and $c_s^2\rightarrow 1/3$, which provide a measure of the vanishing of the stress-energy tensor. In~\cite{Fujimoto:2022ohj}, it was argued that $\Delta$ might monotonically approach zero with increasing energy density, while at the same time $c_s^2$ develops a peak at lower densities. Such description of dense matter is naturally obtained in quarkyonic models~\cite{McLerran:2007qj, Duarte:2021tsx, Kojo:2021ugu, Kojo:2021hqh, Fukushima:2015bda, McLerran:2018hbz, Jeong:2019lhv, Sen:2020peq, Cao:2020byn, Kovensky:2020xif}.

In this work, we demonstrate that matter inside the cores of maximally massive NSs becomes almost conformal. We analyze the properties of the speed-of-sound, trace anomaly, and net-baryon number susceptibility in the context of medium composition and possible critical behavior or its remnants.

\section{Methods}

We construct an ensemble of EoSs based on the piecewise-linear speed-of-sound parametrization introduced in~\cite{Annala:2019puf}. The model has been already used in several other works~\cite{Annala:2021gom, Altiparmak:2022bke, Ecker:2022xxj, Ecker:2022dlg, Jiang:2022tps}. For details, see~\cite{Marczenko:2022jhl}. The EoSs are required to be consistent with the pQCD results down to $\simeq 40n_0$~\cite{Fraga:2013qra}. In addition, we impose the observational astrophysical constraints: the lower bound of the maximum-mass constraint, $M_{\rm TOV}\geq(2.08\pm0.07)M_\odot$, from the measurement of J0740+6620~\cite{Fonseca:2021wxt}, and the constraint on tidal deformability of a $1.4M_\odot$ NS, $\Lambda_{1.4} =190^{+390}_{-120}$~\cite{LIGOScientific:2018cki}, from GW170817 event measured by the LIGO/Virgo Collaboration (LVC). In total, we analyzed a sample of $4.62\times10^5$ EoSs that fulfill the imposed observational and pQCD constraints.

\section{Results and Discussion}
\begin{figure}
    \centering
    \includegraphics[width=.7\linewidth]{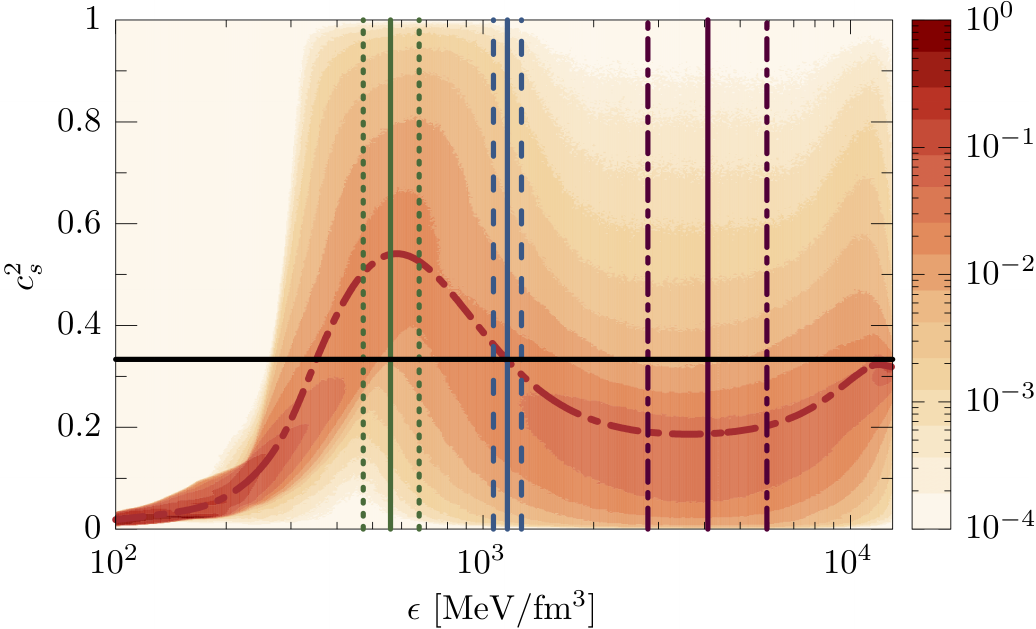}\;\;\;
    \caption{Probability density function (PDF) of the speed of sound as a function of the energy density. The red, dash-dotted lines show the averages of these quantities. Vertical lines show the median and $1\sigma$ credibility region for the position of the peak in $c_s^2$ (green solid and dotted lines), values at the center of maximally massive NSs (blue solid and dashed lines), and the position of the peak in $\hat\chi_B$ (purple solid and dash-dotted lines). The horizontal, black line marks the conformal values of $c_s^2=1/3$.}
    \label{fig:cs2}
\end{figure}

In Fig.~\ref{fig:cs2}, we show the probability distribution function (PDF) of the speed of sound as a function of energy density. The speed of sound swiftly increases at low densities and generates a peak above $1/3$. Notably, the distribution seems to reflect the generic peak-dip structure, and, at large densities, $c_s^2$ converges to the pQCD result. A similar PDF was also obtained in~\cite{Altiparmak:2022bke}. We have located the local peaks in $c_s^2$. We obtain the location of the energy density, $\epsilon_{\rm peak} = 0.559^{+0.110}_{-0.088}~\rm GeV/fm^3$ and net-baryon density $n_{B,\rm peak} = 0.54^{+0.09}_{-0.07}~\rm fm^{-3}$ at the peak position at $1\sigma$ confidence level. We also find the estimate of the median for the largest central energy density found in neutron stars, $\epsilon_{\rm TOV} = 1.163^{+0.107}_{-0.97}~\rm GeV/fm^3$ at $1\sigma$ confidence level. The corresponding value of the speed of sound is $c_{s, \rm TOV}^2 = 0.28\pm0.06$ at $1\sigma$ confidence level, which is remarkably close to the conformal value. We conclude that it is plausible for heavy NSs to feature a peak in the speed of sound in their cores with $c_s^2 > 1/3$ which then decreases to $\sim 1/3$. Such non-monotonicity has been recently conjectured~\cite{Kojo:2020krb} and observed in statistical analysis of the EoS~\cite{Ecker:2022xxj}. For instance, the quarkyonic description of dense matter leads to a rapid increase, accompanied by a peak in the speed of sound~\cite{McLerran:2018hbz, McLerran:2007qj, Hidaka:2008yy}.

\begin{figure}
    \centering
    \includegraphics[width=.7\linewidth]{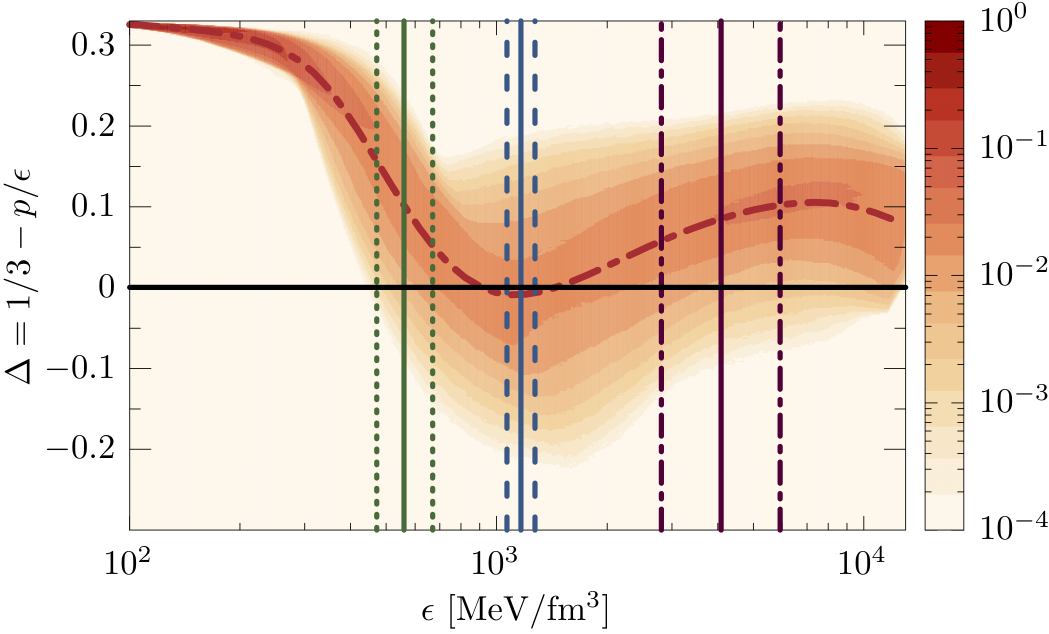}
    \caption{The same as in Fig.~\ref{fig:cs2} but for trace anomaly. The horizontal, black line marks the conformal value of $\Delta = 0$.}
    \label{fig:trace}
\end{figure}

\begin{figure}[t!]
    \centering
    \includegraphics[width=.7\linewidth]{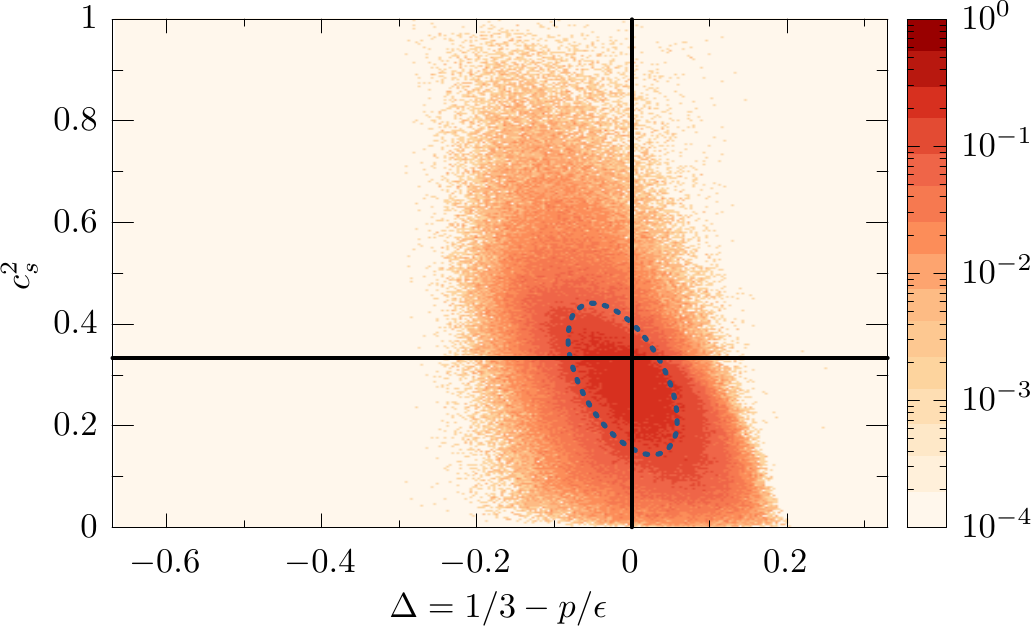}
    \caption{PDF of the speed of sound vs trace anomaly at the center of maximally stable NSs. The dotted, blue ellipse marks the $1\sigma$ credibility around the mean. The black, solid lines mark the conformal values.}
    \label{fig:trace_cs2_maxM}
\end{figure}

In Fig.~\ref{fig:trace}, we plot the calculated PDF of the trace anomaly. It decreases monotonically up to $\epsilon \sim\epsilon_{\rm TOV}$ as shown in~\cite{Fujimoto:2022ohj}. In Fig.~\ref{fig:trace_cs2_maxM}, we plot the PDF of the central values of the speed of sound and trace anomaly obtained for the maximally massive stellar configurations.We obtain values of the median $c_{s, \rm TOV}^2 = 0.28\pm0.06$ and $\Delta_{\rm TOV} = -0.01\pm0.03$ at the $1\sigma$ confidence level. This is remarkably close to the conformal value and suggests the existence of strongly-coupled conformal matter at the cores of massive NSs.

Deconfinement can be phenomenologically linked to the percolation threshold of hadrons of a given size~\cite{Magas:2003wi, Castorina:2008vu, Satz:1998kg, Fukushima:2020cmk, Braun-Munzinger:2014lba}. In the percolation theory of objects with constant volume
\begin{equation}
V_0 = (4/3)\pi R_0^3 \textrm,
\end{equation}
the critical density is given by~\cite{Braun-Munzinger:2014lba}
\begin{equation}
n^{\rm per}_c = 1.22/V_0 \textrm.
\end{equation}
Taking the average proton mass radius $R_0 = 0.80 \pm 0.05$~fm~\cite{Wang:2022uch}, yields $n^{\rm per}_c = 0.57^{+0.12}_{-0.09}~\rm fm^{-3}$, which is consistent with density extracted in this work at which the speed of sound reaches its maximum. Moreover, In Pb-Pb collisions at $\sqrt s=2.76~\rm TeV$ hadrons are produced at the QCD phase boundary at $T_{\rm pc}$ from the fireball of volume $V=4175\pm 380~\rm fm^{3}$~\cite{Andronic:2017pug, Andronic:2018qqt, Braun-Munzinger:2014lba}. Taking the ratio of the number of hadrons per unit of rapidity $N_t=2486\pm 146$, measured by ALICE collaboration, and the above fireball volume, one gets $n_c=0.596\pm 0.065~\rm fm^{-3}$. This value is also consistent with the critical percolation density and the extracted density $n_{B,\rm peak}$ at the peak position of the speed of sound. One can conclude that quark deconfinement could be linked to a non-monotonic behavior of $c_s^2$, thus one can identify the maximum of $c_s^2$ as being due to the change in medium composition, from nuclear to quark or quarkyonic matter.

We note that the behavior of the speed of sound shown in Fig.~\ref{fig:cs2} is very different from that obtained in QCD matter at finite temperature~\cite{Redlich:1985uw, HotQCD:2014kol}, where $c_s^2 < 1/3$ and exhibits a minimum at the critical energy density, $\epsilon_c=0.42\pm 0.06~\rm GeV/fm^{3}$, where chiral symmetry is partially restored and quarks are deconfined~\cite{HotQCD:2018pds}. This is linked to the domination of attractive interactions with resonance formation~\cite{Castorina:2009de}, as opposed to cold nuclear matter, where repulsive interactions are dominant, which implies increasing $c_s^2$~\cite{Zeldovich:1961sbr}.

Lastly, to quantify possible critical signals in the EoS, we consider the dimensionless second-order cumulant of the net-baryon density, $\hat\chi_B \equiv \chi_B /\mu_B^2 = n_B / (\mu_B^3 c_s^2)$, and locate its global maximum, $\hat \chi_B^{\rm max}$. Note that we neglect signals from the liquid-gas phase transition at low densities. We find that more than $93\%$ of the global maxima lie in the density domain beyond the gravitational stability ($\epsilon > \epsilon_{\rm TOV}$). We find the median $\epsilon_{\rm \chi} = 4.084^{+1.834}_{-1.275}~\rm GeV/fm^3$ at $1\sigma$ confidence level. The onset of criticality which is linked to the maximal baryon-density fluctuations in the EoS is therefore unlikely to be found in the cores of NSs.

We would like to stress that given the near-future progress in multi-messenger astronomy and large-scale nuclear experiments, it is necessary to understand the dense matter EoS in terms of fundamental properties of strong interactions, i.e., restoration of chiral symmetry inside NSs, as well as the onset of quark or quarkyonic matter. This can be possibly achieved within the application of advanced parity doublet models with the elusive interplay between chiral symmetry breaking~\cite{Marczenko:2021uaj, Marczenko:2022hyt}, deconfinement~\cite{Benic:2015pia, Marczenko:2017huu}, and the topology at the Fermi surface~\cite{Kojo:2021ugu}. Work in this direction is in progress.

\section*{Acknowledgments}
This work is supported partly by the Polish National Science Centre (NCN) under OPUS Grant No. 2018/31/B/ST2/01663 (K.R. and C.S.), Preludium Grant No. 2017/27/N/ST2/01973 (M.M.), and the program Excellence Initiative–Research University of the University of Wroc\l{}aw of the Ministry of Education and Science (M.M). The work of L. M. was supported by the U.S. DOE under Grant No. DE-FG02-00ER41132. K.R. also acknowledges the support of the Polish Ministry of Science and Higher Education. 
\bibliography{references}

\end{document}